\documentclass[submission,copyright,creativecommons]{eptcs}

\providecommand{\event}{EPTCS}

\usepackage[english]{babel}
\usepackage[utf8x]{inputenc}
\usepackage[T1]{fontenc}
\usepackage{color}

\newcommand{\itogatp}{{\sc i2gatp}}
\newcommand{\TGTP}{\textit{TGTP\/}}
\def\Coq{{\it Coq}}
\def\Isabelle{{\it Isabelle}}

\title{Towards Ranking Geometric Automated Theorem Provers}

\author{Nuno Baeta
  \institute{
    CISUC \\
    University of Coimbra, Portugal} \\
  \email{nmsbaeta@gmail.com} \\
  \and
  Pedro Quaresma
  \institute{
    CISUC / Department of Mathematics \\
    University of Coimbra, Portugal} \\
  \email{pedro@mat.uc.pt}
}

\def\authorrunning{N. Baeta \& P. Quaresma}

\def\titlerunning{Towards Ranking Geometric Automated Theorem Provers}

\begin{document}

\maketitle

% Abstract
\begin{abstract}
  % Sobre os GATPs
  The field of geometric automated theorem provers has a long and rich
  history, from the early AI approaches of the 1960s, synthetic
  provers, to today algebraic and synthetic provers.

  % sobre a área da dedução automática e a especificidade da geometria
  The geometry automated deduction area differs from other areas by
  the strong connection between the axiomatic theories and its
  standard models. In many cases the geometric constructions are used
  to establish the theorems' statements, geometric constructions are,
  in some provers, used to conduct the proof, used as counter-examples
  to close some branches of the automatic proof. Synthetic geometry
  proofs are done using geometric properties, proofs that can have a
  visual counterpart in the supporting geometric construction.  
  
  % Objectivo e sua justificação
  With the growing use of geometry automatic deduction tools as
  applications in other areas, e.g. in education, the need to evaluate
  them, using different criteria, is felt. Establishing a ranking
  among geometric automated theorem provers will be useful for the
  improvement of the current methods/implementations.  Improvements
  could concern wider scope, better efficiency, proof readability and
  proof reliability.

  % Operacionalização do objectivo proposto
  To achieve the goal of being able to compare geometric automated
  theorem provers a common test bench is needed: a common language to
  describe the geometric problems; a comprehensive repository of
  geometric problems and a set of quality measures.
\end{abstract}

% Introduction
\section{Introduction}
\label{sec:introduction}

The first geometric automated theorem provers proposed, that came as
early as in 1959~\cite{Gelertner1959}, adapt general-purpose reasoning
approaches developed in the field of artificial intelligence,
automating the traditional geometric proving processes. In order to
avoid combinatorial explosion while applying postulates, many suitable
heuristics, e.g. adding auxiliary elements to the geometric
configuration, have been developed. Although being able to produce
readable proofs, the proposed methods were very narrow-scoped and not
efficient~\cite{Wu1984}.

Recent results using this approach, like the \emph{ArgoCLP} theorem
prover, based on coherent logic~\cite{Stojanovic2011a}, or the
deductive database approach~\cite{chou2000a} having been proposed with
different rates of success in different classes of geometric problems.

% The 'old' methods - algebraic
The algebraic methods, such as the characteristic set
method~\cite{Chou1985,Wu1984}, the polynomial elimination
method~\cite{Wang95}, the Gr\"obner basis method~\cite{Kapur1986}, and
the Clifford algebra approach~\cite{Li2000}, reduce the complexity of
logical inferences by computing relations between coordinates of
geometric entities. What is gained in efficiency and wider scope is
lost in the connection of the algebraic proof and the geometric
reasoning. These methods are broad-scope, efficient, but if,
eventually, a proof record is produced, it will be a very complex
algebraic proof~\cite{Chou1994b,Wu1984}. This is still an active area
of
research~\cite{Botana2015,Janicic2012a,Janicic2006a,Kovacs2015,Ye2011}.

% The new methods - semi-synthetic/synthetic
In order to combine the readability of synthetic methods and
efficiency of algebraic methods, some approaches, such as the area
method~\cite{Chou1996a,Janicic2012a}, the full angle
method~\cite{Chou1996b} represent geometric knowledge in a form of
expressions with respect to geometric invariants. These methods are
broad-scoped (less than the algebraic), efficient (less than the
algebraic) and capable of producing readable proofs. An implementation
as a \emph{Coq}\footnote{\url{https://coq.inria.fr/}} contribution is
such that it can have theirs proofs
verified~\cite{boutry2014a,Janicic2012a}.

When considering ranking the Geometric Automated Theorem Provers
(GATP) we have to consider some, somehow opposing, goals: scope, the
geometries being considered and what kind of problems are provable;
efficiency, the time needed to complete the proof; readability of the
proof produced (if any); and, with the help of proof assistants, the
reliability of generated proofs~\cite{Jiang2012,Quaresma2017a}.

To be able to compare the different methods and implementations, a
test bench must be defined, a common language to state the geometric
theorems, a comprehensive repository of geometric problems and a set
of measures of quality capable of assessing the GATPs in different
classes, such as: scope; efficiency; readability; reliability of
generated proofs.

\textit{Overview of the paper.}  The paper is organised as follows:
first, in Section~\ref{sec:testBench}, the test bench is discussed.
In Section~\ref{sec:rankingGATP}, the different measures of quality
are discussed. Final conclusions are drawn and future work is foreseen
in Section~\ref{sec:conclusionFutureWork}.

% Operacionalização do objectivo proposto
\section{Test Bench}
\label{sec:testBench}

% TGTP como bancada de trabalho 

In order to implement a test bench, the Thousands of Geometric
problems for geometric Theorem
Provers\footnote{\url{http://hilbert.mat.uc.pt/TGTP/}} (\TGTP)
platform presents itself as a solid foundation to fulfil such purpose.
{\TGTP} aims to provide the automated reasoning in geometry community
with a comprehensive and easily accessible, library of GATP test
problems~\cite{Quaresma2011}, it already provides a centralised common
repository of geometric problems\footnote{As of 2018--04--21 there
  were 236 problems.}, an unambiguous reference mechanism, textual and
geometric search mechanism and the problems are kept in the {\itogatp}
common format~\cite{Quaresma2015a}. Moreover, \TGTP\ provides, as part
of its infra\-structure, implementations of several methods, namely:
\emph{GCLC}\footnote{\url{http://poincare.matf.bg.ac.rs/~janicic/gclc/}}
implementations of Wu's method, Gr\"obner basis method and the area
method; and a \emph{Coq} implementation of the area
method\footnote{\url{https://github.com/coq-contribs/area-method}}~\cite{NarbouxContribArea}.
Statistical and performance information is also supplied for all
implementations, as well as a proof status for each geometric problem.

With a working test bench in operation, an interesting goal to pursue
would be to operationalize a competition between GATPs, similar to
\emph{CASC}~\cite{Sutcliffe2018}.  Its ultimate goal, like in
\emph{CASC}, would be to encourage researchers to improve existing
GATPs and implement new ones.

For example, in table~\ref{tab:comparingGATP} the CPU
times\footnote{Linux 4.9.0-2-amd64 \#1 SMP Debian 4.9.18-1
  (2017-03-30), Intel(R) Core(TM) i7-4770 CPU @ 3.40GHz. The used
  ``time-out'' threshold was 60s, but it can be configured for any
  given value.} needed to complete the proofs for some {\TGTP}
problems are presented. The GATPs used were: \emph{Coq};\footnote{For
  some of the examples it was not possible to have the transcription
  of the problem in \emph{Coq}} \emph{GCLC}; \emph{GeoGebra};
\emph{OpenGeoProver} (OGP) with the methods: \emph{area method} (AM);
\emph{Wu's method} (WM); \emph{Gr{\"o}bner Basis' method} (GBM);
\emph{BotanaGiac} (BG)~\cite{Botana2015a}. The \emph{OGP} and
\emph{GCLC} implementations of the Wu's method stand-off as the only
implementations that were able to prove all the conjectures, with
\emph{GCLC} having a marginal advantages over \emph{OGP}.

\begin{table}[hbtp]
  \centering
  \begin{tabular}{ccccccc}
    & \textbf{Coq} & \multicolumn{3}{c}{\textbf{GCLC}} &
                       \textbf{GeoGebra} & \textbf{OGP} \\ \cline{3-5}
    & AM       & AM       & WM    & GBM      & BG & WM \\ \hline
    GEO0080 & 0.74     & 0.012    & 0.016 & time-out & 0.615      & 0.028 \\
    GEO0226 & ---      & 0.748    & 1.62  & time-out & time-out   & 0.075 \\
    GEO0228 & ---      & 0.008    & 0.012 & 0.036    & 0.207      & 0.015 \\
    GEO0237 & time-out & time-out & 0.048 & time-out & 0.264      & 0.105 \\
    GEO0238 & time-out & 0.032    & 0.024 & 0.092    & 0.18       & 0.102 \\
    GEO0240 & ---      & time-out & 0.056 & time-out & 1.314      & 0.030 \\ \hline
  \end{tabular}
  \caption{Comparing GATPs}
  \label{tab:comparingGATP}
\end{table}

But, apart CPU times, when ranking GATPs other issued must be
considered: the scope, the proof readability and proof
reliability.

The \emph{GCLC} implementation of the area method is the only GATP
providing a readable synthetic geometric proof.

The \emph{GCLC} implementation of the Wu's method is the one with the
wider scope (for this small set of examples), being able to prove all
the problems.

The \emph{Coq} implementation of the area method is the only GATP
formally verified~\cite{boutry2014a,Janicic2012a}. The \emph{Coq}
proof assistant relies on a small kernel, extensively verified,
against whom all the other layers are formally checked. The \emph{Coq}
implementation of the area method is thus formally verified. For all
the other GATPs the implementations trustworthiness relies on the
GATPs code as a whole, without any formal verification of its
correctness. This issue is more important when the algebraic
implementations are to be considered, in those cases its black box
nature, with only a yes/no answer, prevent the user to manually check
the proof produced. Any new implementation should be extensively
checked before it can be considered finished.

The intended use is also important. For example, in education two
uses of a GATP are foreseen, the validation of a given conjecture and/or
construction~\cite{Botana2015a,Janicic2007a} and the proof itself as
object of study.

For the first case (validation), the proof is not needed, only a
formal validation is needed, speed is the most important measure of
quality. Rowe~\cite{Rowe1972} and, later, Stahl~\cite{Stahl1994}
studied this issue, defining the ``Post-Teacher Question Wait-Time'',
i.e. the time a high-school class will wait in silence after a teacher
place a question. This is a (surprisingly!?) short span of time, a
good response time is less then 1.5s, a fair response time will be
less then 3s, everything above that would be unsuitable for a
high-school classroom. Looking in table~\ref{tab:comparingGATP}, we
can see that the implementations of the algebraic methods (with the
exception of \emph{GCLC} implementation of the Gr\"obner basis) gave
good results for all but two cases. One result in the border between
good and fair and a ``time-out''.

For the second case a readable proof is needed, i.e. a synthetic or
semi-synthetic geometric proof readable, by teachers (at least) and
students. For the first case the algebraic methods seems to be the
best one, for the second case they are completely useless, the proofs
produced are algebraic, unrelated with the geometric construction in
consideration. The synthetic or even semi-synthetic should be
considered instead.
 
As far as the authors of this paper know, there are two proposals to
measure the readability of a proof: Chou et al.~\cite{chou1994a}
proposed a way to measure how difficult a formal proof is (using the
area method) based on the time and the number of steps needed to
perform it; de~Bruijn proposed the de~Bruijn
factor~\cite{debruijn1994a,Wiedijk2000},\footnote{\url{https://www.cs.ru.nl/~freek/factor/}}
the quotient of the \emph{size} of corresponding informal proof and
the \emph{size} of the formal proof. Apart this quantitative measures
some qualitative considerations and also considerations about the
audience in consideration led to a schema, proposed
in~\cite{Quaresma2019}, to classify the proofs produced by GATPs:

\begin{enumerate}
\item no readable proof;
\item non-synthetic proof (i.e. a proof without a correspondent
  geometric description, e.g. algebraic methods);
\item semi-synthetic proof with a corresponding prover's language
  rendering; \label{sspl}
\item (semi-)synthetic proof with a corresponding natural language
  rendering; \label{ssnl}
\item (semi-)synthetic proof with a corresponding natural language and
  visual renderings~\cite{ye2010a,ye2010b}. \label{ssnlv}
\end{enumerate}

The synthetic proofs with a corresponding natural language and visual
renderings is the most difficult to attain, but also the more
desirable, e.g. in an educational setting.

% Ratings
\section{Ranking GATP}
\label{sec:rankingGATP}

To be able to compare the different methods and implementations,
alongside a standard test bench (see Section~\ref{sec:testBench}), we
must define a set of measures of quality capable of assessing the
GATPs in different classes: scope; efficiency; readability;
reliability of generated proofs.

\begin{description}
\item[Scope] To measure the scope of a GATP, one should consider:
  \begin{itemize}
  \item which geometries are allowed by the GATP---despite the fact that
    most GATPs deal only with Euclidean geometry, some exist that prove
    geometric problems in non-Euclidean
    geometries~\cite{chou1988a,zhang1995a};
  \item what kind of problems are provable---as an example, the area
    method uses \emph{geometric invariants} as basic quantities to prove
    theorems, each of which is used to deal with different geometric
    relations, hence allowing certain theorems to be easily
    proved~\cite{chou1994a}.
  \end{itemize}

  Considering the existing methods, algebraic ones have the broadest
  scope.  Not only have they been used to prove theorems in Euclidean
  and non-Euclidean geometry, but, for every geometry, the range of
  difficulty of the problems proved is very wide~\cite{chou1988a}.
  The earlier synthetic approaches were narrow scoped, e.g. the
  \emph{GEOM}~\cite{coelho1986a} only dealt with a limited set of
  geometric elements and relations~\cite{chou1994a}. In between lie
  semi-synthetic methods and coherent logic based methods.
  Semi-synthetic methods, starting with the area method which is
  complete for \emph{constructive
    geometry}~\cite{Chou1996a,Chou1996b,Janicic2012a}, with its
  \emph{geometric invariant}, the \emph{signed area}, allow many
  problems with relations like incidence and parallelism to be proved.
  Adding another \emph{geometric invariant}, the \emph{Pythagoras
    difference}, allows the problems with relations like
  perpendicularity and congruence of line segments to be easily
  proved.  Adding other~\emph{geometric invariants} such as the
  \emph{full-angle} (which gives its name to the full-angle method),
  the \emph{volume} and the \emph{vector}, allows the demonstration of
  an ever increasing range of theorems~\cite{chou1994a}.

  Synthetic and semi-synthetic methods scope may be influenced by the use
  of deductive data\-bases~\cite{chou2000a,coelho1986a}.  Indeed, as
  stated in \cite{chou2000a}, unexpected results may be obtained, some
  of which are possibly new.

\item[Efficiency] The purpose of a GATP is, in addition to prove
  geometric conjectures, that these are obtained efficiently. By and
  large, although other resources may be involved, efficiency is related
  to time and memory space: we look for algorithms/implementations capable of
  fulfilling a proof in a reasonable amount of time and space.

  Time is indeed the \emph{natural} way to measure efficiency since it
  is used extensively, if not exclusively, throughout the
  literature~\cite{chou1988a,Chou1994b,Chou1996a,Chou1996b,chou2000a,Janicic2012a}.
  and, for obvious reasons, in a competition such as \emph{CADE ATP
    System Competition} (\emph{CASC}).  Moreover, such measure is of
  paramount importance when considering an educational environment.

  Note however that when authors state something about GATP times,
  they do so in their settings, i.e. computer and operating system
  used, somehow restricting the usefulness of these results.  The
  existence of a free and open platform where different GATPs can be
  tested on equal terms proves to be of utmost importance. In
  table~\ref{tab:comparingGATP} the examples were tested under the
  same system being comparable, an ongoing work will allow using all
  the conjectures in {\TGTP} to test all those GATPs under the same
  system. 

  Regarding space, the authors are unaware of any study, presumably
  because these physical constraints are nowadays less important.
  Besides, from the users point of view time is the most important
  factor.

\item[Readability] Until recently, proofs in mathematics were solely
  made and verified by humans.  With the advent of computers and
  automated reasoning that is no longer the case.  As in some settings
  readability of a proof (by a human) is of utmost importance, such
  characteristic is considered crucial. Indeed, in an educational
  setting the proof is an object of learning by itself and the ability
  of a GATP to produce a synthetic proof, with the usual geometric
  inference rules, is of fundamental importance to its usability.

  {\TGTP} both the quantitative and qualitative measures should be
  introduce in such a way that a user can filter the information in
  the way that best fit his/her needs.
  
\item[Reliability] By reliability is meant the confidence that we
  have in the proofs made by a given prover. Is the prover correctly
  implemented? 

  Using proof assistants like {\Coq}, or
  {\Isabelle},\footnote{\url{https://isabelle.in.tum.de/}} the
  implementation of a given prover can be formally established,
  e.g. in~\cite{NarbouxContribArea} a implementation of the area
  method within {\Coq} is described, where all the properties of the
  geometric quantities required by the area method are verified,
  demonstrating the correctness of the system, reducing concerns of
  reliability, to the trustworthiness of respective proof
  assistants~\cite{Janicic2012a,NarbouxContribArea}.

  For the other implementations, the trustworthiness of GATP relies on
  the implementation code as a whole and in the extensive testing of
  the new implementations, e.g. using a workbench like \emph{TGTP}.
\end{description}

% Conclusion and Future Work
\section{Conclusion and Future Work}
\label{sec:conclusionFutureWork}

When regarding the area of geometric automated deduction the
evaluation of automated provers is no longer a mono-objective problem,
time considerations are important but scope and readability of the
proofs are also very important.

Maybe the question ``what is the best GATP of them all'', can not be
answered, but at least we should have some partials answers when
looking for a GATP that fit a particular goal.

The intended audience is also a point to be considered, e.g the narrow
scope provers with specific heuristics, closing the circle with the
early AI approaches of the 1960s, may be not relevant to an expert in
the field of automated theorem proving, but the definition of
sub-languages for educational use, could led to narrow scope GATPs
with good qualities when readability of the proofs is to be
considered, in specific educational settings.

We need a common workbench: a common language to describe the
geometric problems; a large set of problems and a set of
measures. Building on {\TGTP} test bench~\cite{Quaresma2011},
{\itogatp} common language~\cite{Quaresma2015a} and \emph{CASC}, CADE
ATP System Competition~\cite{Sutcliffe2018}, the first two need to be
improved and better integrated, and adapting the ideas of the last, a
common test bench should be built, with the goal of helping GATPs
developers to improve their systems, to better fit the needs of the
different communities of users. 

% Bibliography
\bibliographystyle{eptcs}

%\bibliography{geotiles}

\begin{thebibliography}{10}
\providecommand{\bibitemdeclare}[2]{}
\providecommand{\surnamestart}{}
\providecommand{\surnameend}{}
\providecommand{\urlprefix}{Available at }
\providecommand{\url}[1]{\texttt{#1}}
\providecommand{\href}[2]{\texttt{#2}}
\providecommand{\urlalt}[2]{\href{#1}{#2}}
\providecommand{\doi}[1]{doi:\urlalt{http://dx.doi.org/#1}{#1}}
\providecommand{\bibinfo}[2]{#2}

\bibitemdeclare{article}{Botana2015a}
\bibitem{Botana2015a}
\bibinfo{author}{Francisco \surnamestart Botana\surnameend},
  \bibinfo{author}{Markus \surnamestart Hohenwarter\surnameend},
  \bibinfo{author}{Predrag \surnamestart Jani{\v{c}}i{\'c}\surnameend},
  \bibinfo{author}{Zolt{\'a}n \surnamestart Kov{\'a}cs\surnameend},
  \bibinfo{author}{Ivan \surnamestart Petrovi{\'c}\surnameend},
  \bibinfo{author}{Tom{\'a}s \surnamestart Recio\surnameend} \&
  \bibinfo{author}{Simon \surnamestart Weitzhofer\surnameend}
  (\bibinfo{year}{2015}): \emph{\bibinfo{title}{Automated {T}heorem {P}roving
  in {G}eo{G}ebra: {C}urrent {A}chievements}}.
\newblock {\sl \bibinfo{journal}{Journal of Automated Reasoning}}
  \bibinfo{volume}{55}(\bibinfo{number}{1}), pp. \bibinfo{pages}{39--59},
  \doi{10.1007/s10817-015-9326-4}.

\bibitemdeclare{proceedings}{Botana2015}
\bibitem{Botana2015}
\bibinfo{editor}{Francisco \surnamestart Botana\surnameend} \&
  \bibinfo{editor}{Pedro \surnamestart Quaresma\surnameend}, editors
  (\bibinfo{year}{2015}): \emph{\bibinfo{title}{Automated Deduction in
  Geometry, 10th International Workshop, ADG 2014}}. {\sl
  \bibinfo{series}{Lecture Notes in Artificial Intelligence}}
  \bibinfo{volume}{9201}, \bibinfo{publisher}{Springer}.
  \doi{10.1007/978-3-319-21362-0}.

  
\bibitemdeclare{inproceedings}{boutry2014a}
\bibitem{boutry2014a}
\bibinfo{author}{Pierre \surnamestart Boutry\surnameend},
  \bibinfo{author}{Julien \surnamestart Narboux\surnameend},
  \bibinfo{author}{Pascal \surnamestart Schreck\surnameend} \&
  \bibinfo{author}{Gabriel \surnamestart Braun\surnameend}
  (\bibinfo{year}{2014}): \emph{\bibinfo{title}{Using small scale automation to
  improve both accessibility and readability of formal proofs in geometry}}.
\newblock In \bibinfo{editor}{Francisco \surnamestart Botana\surnameend} \&
  \bibinfo{editor}{Pedro \surnamestart Quaresma\surnameend}, editors: {\sl
  \bibinfo{booktitle}{Preliminary Proceedings of the 10th International
  Workshop on Automated Deduction in Geometry, ADG~2014, Coimbra, Portugal,
  9--11~July, 2014}}, {\sl \bibinfo{series}{CISUC Tech Reports}}
  \bibinfo{volume}{TR 2014/01}, pp. \bibinfo{pages}{31--49}.
\newblock
  \urlprefix\url{https://www.cisuc.uc.pt/ckfinder/userfiles/files/TR 2014-01.pdf}.

\bibitemdeclare{inbook}{debruijn1994a}
\bibitem{debruijn1994a}
\bibinfo{author}{Nicolaas~Govert \surnamestart de~Bruijn\surnameend}
  (\bibinfo{year}{1994}): \emph{\bibinfo{title}{A {S}urvey of the {P}roject
  {A}utomath}}, chapter \bibinfo{chapter}{A {S}urvey of the {P}roject
  {A}utomath}, pp. \bibinfo{pages}{141--161}.
\newblock {\sl \bibinfo{series}{Studies in Logic and the Foundations of
  Mathematics}} \bibinfo{volume}{133}, \bibinfo{publisher}{North Holland},
  \doi{10.1016/S0049-237X(08)70203-9}.
%\newblock \urlprefix\url{https://pure.tue.nl/ws/files/1892191/597622.pdf}.

\bibitemdeclare{phdthesis}{Chou1985}
\bibitem{Chou1985}
\bibinfo{author}{S.C. \surnamestart Chou\surnameend} (\bibinfo{year}{1985}):
  \emph{\bibinfo{title}{Proving and discovering geometry theorems using {W}u's
  method}}.
\newblock Ph.D. thesis, \bibinfo{school}{The University of Texas, Austin}.

\bibitemdeclare{book}{chou1988a}
\bibitem{chou1988a}
\bibinfo{author}{Shang-Ching \surnamestart Chou\surnameend}
  (\bibinfo{year}{1988}): \emph{\bibinfo{title}{Mechanical Geometry Theorem
  Proving}}.
\newblock {\sl \bibinfo{series}{Mathematics and Its
  Applications}}~\bibinfo{volume}{41}, \bibinfo{publisher}{D.~Reidel Publishing
  Company}.

\bibitemdeclare{techreport}{Chou1994b}
\bibitem{Chou1994b}
\bibinfo{author}{Shang-Ching \surnamestart Chou\surnameend},
  \bibinfo{author}{Xiao-Shan \surnamestart Gao\surnameend} \&
  \bibinfo{author}{Jing-Zhong \surnamestart Zhang\surnameend}
  (\bibinfo{year}{1994}): \emph{\bibinfo{title}{A {C}ollection of 110
  {G}eometry {T}heorems and {T}heir {M}achine {P}roduced {P}roofs {U}sing
  {F}ull-{A}ngles}}.
\newblock \bibinfo{type}{Technical Report} \bibinfo{number}{TR-94-4},
  \bibinfo{institution}{Department of Computer Science, Wichita State
  University}.
\newblock \urlprefix\url{https://www.researchgate.net/publication/239564904}.

\bibitemdeclare{book}{chou1994a}
\bibitem{chou1994a}
\bibinfo{author}{Shang-Ching \surnamestart Chou\surnameend},
  \bibinfo{author}{Xiao-Shan \surnamestart Gao\surnameend} \&
  \bibinfo{author}{Jing-Zhong \surnamestart Zhang\surnameend}
  (\bibinfo{year}{1994}): \emph{\bibinfo{title}{Machine Proofs in Geometry:
  Automated Production of Readable Proofs for Geometry Problems}}.
\newblock {\sl \bibinfo{series}{Applied Mathematics}}~\bibinfo{volume}{6},
  \bibinfo{publisher}{World Scientific}, \doi{10.1142/9789812798152}.
\newblock \urlprefix\url{https://www.researchgate.net/publication/240102887}.

\bibitemdeclare{article}{Chou1996a}
\bibitem{Chou1996a}
\bibinfo{author}{Shang-Ching \surnamestart Chou\surnameend},
  \bibinfo{author}{Xiao-Shan \surnamestart Gao\surnameend} \&
  \bibinfo{author}{Jing-Zhong \surnamestart Zhang\surnameend}
  (\bibinfo{year}{1996}): \emph{\bibinfo{title}{Automated {G}eneration of
  {R}eadable {P}roofs with {G}eometric {I}nvariants: {I}. {M}ultiple and
  {S}hortest {P}roof {G}eneration}}.
\newblock {\sl \bibinfo{journal}{Journal of Automated Reasoning}}
  \bibinfo{volume}{17}(\bibinfo{number}{3}), pp. \bibinfo{pages}{325--347},
  \doi{10.1007/bf00283133}.

\bibitemdeclare{article}{Chou1996b}
\bibitem{Chou1996b}
\bibinfo{author}{Shang-Ching \surnamestart Chou\surnameend},
  \bibinfo{author}{Xiao-Shan \surnamestart Gao\surnameend} \&
  \bibinfo{author}{Jing-Zhong \surnamestart Zhang\surnameend}
  (\bibinfo{year}{1996}): \emph{\bibinfo{title}{Automated {G}eneration of
  {R}eadable {P}roofs with {G}eometric {I}nvariants: {II}. {T}heorem {P}roving
  {W}ith {F}ull-{A}ngles}}.
\newblock {\sl \bibinfo{journal}{Journal of Automated Reasoning}}
  \bibinfo{volume}{17}(\bibinfo{number}{3}), pp. \bibinfo{pages}{349--370},
  \doi{10.1007/BF00283134}.

\bibitemdeclare{article}{chou2000a}
\bibitem{chou2000a}
\bibinfo{author}{Shang-Ching \surnamestart Chou\surnameend},
  \bibinfo{author}{Xiao-Shan \surnamestart Gao\surnameend} \&
  \bibinfo{author}{Jing-Zhong \surnamestart Zhang\surnameend}
  (\bibinfo{year}{2000}): \emph{\bibinfo{title}{A {D}eductive {D}atabase
  {A}pproach to {A}utomated {G}eometry {T}heorem {P}roving and {D}iscovering}}.
\newblock {\sl \bibinfo{journal}{Journal of Automated Reasoning}}
  \bibinfo{volume}{25}(\bibinfo{number}{3}), pp. \bibinfo{pages}{219--246},
  \doi{10.1023/A:1006171315513}.

\bibitemdeclare{article}{coelho1986a}
\bibitem{coelho1986a}
\bibinfo{author}{Helder \surnamestart Coelho\surnameend} \&
  \bibinfo{author}{Luis~Moniz \surnamestart Pereira\surnameend}
  (\bibinfo{year}{1986}): \emph{\bibinfo{title}{Automated {R}easoning in
  {G}eometry {T}heorem {P}roving with {P}rolog}}.
\newblock {\sl \bibinfo{journal}{Journal of Automated Reasoning}}
  \bibinfo{volume}{2}(\bibinfo{number}{4}), pp. \bibinfo{pages}{329--390},
  \doi{10.1007/BF00248249}.

\bibitemdeclare{inproceedings}{Gelertner1959}
\bibitem{Gelertner1959}
\bibinfo{author}{H.~\surnamestart Gelernter\surnameend} (\bibinfo{year}{1995}):
  \emph{\bibinfo{title}{Realization of a geometry-theorem proving machine}}.
\newblock In: {\sl \bibinfo{booktitle}{Computers \& thought}},
  \bibinfo{publisher}{MIT Press}, \bibinfo{address}{Cambridge, MA, USA}, pp.
  \bibinfo{pages}{134--152}, \bibinfo{isbn}{ISBN: 0-262-56092-5}.
  
\bibitemdeclare{article}{Janicic2012a}
\bibitem{Janicic2012a}
\bibinfo{author}{Predrag \surnamestart Jani{\v{c}}i{\'c}\surnameend},
  \bibinfo{author}{Julien \surnamestart Narboux\surnameend} \&
  \bibinfo{author}{Pedro \surnamestart Quaresma\surnameend}
  (\bibinfo{year}{2012}): \emph{\bibinfo{title}{The {A}rea {M}ethod: {A}
  {R}ecapitulation}}.
\newblock {\sl \bibinfo{journal}{Journal of Automated Reasoning}}
  \bibinfo{volume}{48}(\bibinfo{number}{4}), pp. \bibinfo{pages}{489--532},
  \doi{10.1007/s10817-010-9209-7}.

\bibitemdeclare{inproceedings}{Janicic2006a}
\bibitem{Janicic2006a}
\bibinfo{author}{Predrag \surnamestart Jani{\v{c}}i{\'c}\surnameend} \&
  \bibinfo{author}{Pedro \surnamestart Quaresma\surnameend}
  (\bibinfo{year}{2006}): \emph{\bibinfo{title}{System {D}escription:
  {GCLC}prover + {G}eo{T}hms}}.
\newblock In \bibinfo{editor}{Ulrich \surnamestart Furbach\surnameend} \&
  \bibinfo{editor}{Natarajan \surnamestart Shankar\surnameend}, editors: {\sl
  \bibinfo{booktitle}{Automated Reasoning: Third International Joint
  Conference, IJCAR~2006, Seattle, WA, USA, August~17--20, 2006, Proceedings}},
  {\sl \bibinfo{series}{Lecture Notes in Artificial Intelligence}}
  \bibinfo{volume}{4130}, \bibinfo{publisher}{Springer}, pp.
  \bibinfo{pages}{145--150}, \doi{10.1007/11814771\_13}.

\bibitemdeclare{inproceedings}{Janicic2007a}
\bibitem{Janicic2007a}
\bibinfo{author}{Predrag \surnamestart Jani{\v{c}}i{\'c}\surnameend} \&
  \bibinfo{author}{Pedro \surnamestart Quaresma\surnameend}
  (\bibinfo{year}{2007}): \emph{\bibinfo{title}{Automatic {V}erification of
  {R}egular {C}onstructions in {D}ynamic {G}eometry {S}ystems}}.
\newblock In \bibinfo{editor}{Francisco \surnamestart Botana\surnameend} \&
  \bibinfo{editor}{Tom{\'a}s \surnamestart Recio\surnameend}, editors: {\sl
  \bibinfo{booktitle}{Automated Deduction in Geometry: 6th International
  Workshop, ADG~2006, Pontevedra, Spain, August~31--September~2, 2006, Revised
  Papers}}, {\sl \bibinfo{series}{Lecture Notes in Artificial Intelligence}}
  \bibinfo{volume}{4869}, \bibinfo{publisher}{Springer}, pp.
  \bibinfo{pages}{39--51}, \doi{10.1007/978-3-540-77356-6\_3}.

\bibitemdeclare{article}{Jiang2012}
\bibitem{Jiang2012}
\bibinfo{author}{Jianguo \surnamestart Jiang\surnameend} \&
  \bibinfo{author}{Jingzhong \surnamestart Zhang\surnameend}
  (\bibinfo{year}{2012}): \emph{\bibinfo{title}{A review and prospect of
  readable machine proofs for geometry theorems}}.
\newblock {\sl \bibinfo{journal}{Journal of Systems Science and Complexity}}
  \bibinfo{volume}{25}(\bibinfo{number}{4}), pp. \bibinfo{pages}{802--820},
  \doi{10.1007/s11424-012-2048-3}.

\bibitemdeclare{article}{Kapur1986}
\bibitem{Kapur1986}
\bibinfo{author}{Deepak \surnamestart Kapur\surnameend} (\bibinfo{year}{1986}):
  \emph{\bibinfo{title}{Using {G}r\"obner bases to reason about geometry
  problems}}.
\newblock {\sl \bibinfo{journal}{Journal of Symbolic Computation}}
  \bibinfo{volume}{2}(\bibinfo{number}{4}), pp. \bibinfo{pages}{399--408},
  \doi{10.1016/S0747-7171(86)80007-4}.

\bibitemdeclare{inbook}{Kovacs2015}
\bibitem{Kovacs2015}
\bibinfo{author}{Zolt{\'a}n \surnamestart Kov{\'a}cs\surnameend}
  (\bibinfo{year}{2015}): \emph{\bibinfo{title}{The Relation Tool in GeoGebra
  5}}, pp. \bibinfo{pages}{53--71}.
\newblock {\sl \bibinfo{series}{Lecture Notes in Artificial Intelligence}}
  \bibinfo{volume}{9201}, \bibinfo{publisher}{Springer International
  Publishing}, \doi{10.1007/978-3-319-21362-0\_4}.

\bibitemdeclare{inproceedings}{Li2000}
\bibitem{Li2000}
\bibinfo{author}{H.~\surnamestart Li\surnameend} (\bibinfo{year}{2000}):
  \emph{\bibinfo{title}{Clifford algebra approaches to mechanical geometry
  theorem proving}}.
\newblock In \bibinfo{editor}{X.-S. \surnamestart Gao\surnameend} \&
  \bibinfo{editor}{D.~\surnamestart Wang\surnameend}, editors: {\sl
  \bibinfo{booktitle}{Mathematics Mechanization and Applications}},
  \bibinfo{publisher}{Academic Press}, \bibinfo{address}{San Diego, CA}, pp.
  \bibinfo{pages}{205--299}, \doi{10.1016/B978-012734760-8/50009-0}.

\bibitemdeclare{misc}{NarbouxContribArea}
\bibitem{NarbouxContribArea}
\bibinfo{author}{Julien \surnamestart Narboux\surnameend}
  (\bibinfo{year}{2009}): \emph{\bibinfo{title}{Formalization of the Area
  Method}}.
\newblock \bibinfo{howpublished}{Coq user contribution}.
\newblock
  \bibinfo{note}{\url{http://dpt-info.u-strasbg.fr/~narboux/area_method.html}}.

\bibitemdeclare{inproceedings}{Paneque2016}
\bibitem{Paneque2016}
\bibinfo{author}{Juan \surnamestart Paneque\surnameend}, \bibinfo{author}{Pedro
  \surnamestart Cobo\surnameend}, \bibinfo{author}{Josep \surnamestart
  Fortuny\surnameend} \& \bibinfo{author}{Philippe \surnamestart {R.
  Richard}\surnameend} (\bibinfo{year}{2016}):
  \emph{\bibinfo{title}{Argumentative Effects of a Geometric Construction
  Tutorial System in Solving Problems of Proof}}.
\newblock In: {\sl \bibinfo{booktitle}{Proceedings of the 4th International
  Workshop on Theorem proving components for Educational software July 15, 2015
  Washington, D.C., USA}}, {\sl \bibinfo{series}{CISUC Technical Reports}}
  \bibinfo{volume}{2016-001}, pp. \bibinfo{pages}{13--35}.
  \newblock
  \urlprefix\url{https://www.cisuc.uc.pt/ckfinder/userfiles/files/TR 2016-01.pdf}.

\bibitemdeclare{incollection}{Quaresma2011}
\bibitem{Quaresma2011}
\bibinfo{author}{Pedro \surnamestart Quaresma\surnameend}
  (\bibinfo{year}{2011}): \emph{\bibinfo{title}{{T}housands of {G}eometric
  Problems for Geometric {T}heorem {P}rovers ({TGTP})}}.
\newblock In \bibinfo{editor}{Pascal \surnamestart Schreck\surnameend},
  \bibinfo{editor}{Julien \surnamestart Narboux\surnameend} \&
  \bibinfo{editor}{J{\"u}rgen \surnamestart Richter-Gebert\surnameend},
  editors: {\sl \bibinfo{booktitle}{Automated Deduction in Geometry}}, {\sl
  \bibinfo{series}{Lecture Notes in Computer Science}} \bibinfo{volume}{6877},
  \bibinfo{publisher}{Springer}, pp. \bibinfo{pages}{169--181},
  \doi{10.1007/978-3-642-25070-5\_10}.

\bibitemdeclare{article}{Quaresma2017a}
\bibitem{Quaresma2017a}
\bibinfo{author}{Pedro \surnamestart Quaresma\surnameend}
  (\bibinfo{year}{2017}): \emph{\bibinfo{title}{Towards an {I}ntelligent and
  {D}ynamic {G}eometry {B}ook}}.
\newblock {\sl \bibinfo{journal}{Mathematics in Computer Science}}
  \bibinfo{volume}{11}(\bibinfo{number}{3--4}), pp. \bibinfo{pages}{427--437},
  \doi{10.1007/s11786-017-0302-8}.

\bibitemdeclare{inproceedings}{Quaresma2015a}
\bibitem{Quaresma2015a}
\bibinfo{author}{Pedro \surnamestart Quaresma\surnameend} \&
  \bibinfo{author}{Nuno \surnamestart Baeta\surnameend} (\bibinfo{year}{2015}):
  \emph{\bibinfo{title}{Current {S}tatus of the {I2GATP} {C}ommon {F}ormat}}.
\newblock In \bibinfo{editor}{Francisco \surnamestart Botana\surnameend} \&
  \bibinfo{editor}{Pedro \surnamestart Quaresma\surnameend}, editors: {\sl
  \bibinfo{booktitle}{Automated Deduction in Geometry: 10th International
  Workshop, ADG~2014, Coimbra, Portugal, July~9--1, 2014, Revised Selected
  Papers}}, {\sl \bibinfo{series}{Lecture Notes in Artificial Intelligence}}
  \bibinfo{volume}{9201}, \bibinfo{publisher}{Springer}, pp.
  \bibinfo{pages}{119--128}, \doi{10.1007/978-3-319-21362-0\_8}.

\bibitemdeclare{article}{Quaresma2019}
\bibitem{Quaresma2019}
\bibinfo{author}{Pedro \surnamestart Quaresma\surnameend},
  \bibinfo{author}{Vanda \surnamestart Santos\surnameend},
  \bibinfo{author}{Pierluigi \surnamestart Graziani\surnameend} \&
  \bibinfo{author}{Nuno \surnamestart Baeta\surnameend} (\bibinfo{year}{2019}):
  \emph{\bibinfo{title}{Taxonomies of geometric problems}}.
\newblock {\sl \bibinfo{journal}{Journal of Symbolic Computation}},
\bibinfo{note}{(in press)}, \doi{10.1016/j.jsc.2018.12.004}.

\bibitemdeclare{article}{Richard2009}
\bibitem{Richard2009}
\bibinfo{author}{Philippe \surnamestart Richard\surnameend},
  \bibinfo{author}{Pedro \surnamestart Cobo\surnameend}, \bibinfo{author}{Josep
  \surnamestart Fortuny\surnameend} \& \bibinfo{author}{Markus \surnamestart
  Hohenwarter\surnameend} (\bibinfo{year}{2009}):
  \emph{\bibinfo{title}{Training teachers to manage problem-solving classes
  with computer support}}.
\newblock {\sl \bibinfo{journal}{Revista de Informática Aplicada / Journal of
  Applied Computing}} \bibinfo{volume}{5}(\bibinfo{number}{1}), pp.
  \bibinfo{pages}{38--50}, \doi{10.13037/rasvol5n1}.
\bibitemdeclare{techreport}{Rowe1972}
\bibitem{Rowe1972}
\bibinfo{author}{Mary~Budd \surnamestart Rowe\surnameend}
  (\bibinfo{year}{1972}): \emph{\bibinfo{title}{Wait-Time and Rewards as
  Instructional Variables: Their Influence on Language, Logic, and Fate
  Control}}.
\newblock \bibinfo{type}{Technical Report}, \bibinfo{institution}{National
  Association for Research in Science Teaching}.
\newblock \urlprefix\url{https://files.eric.ed.gov/fulltext/ED061103.pdf}.

\bibitemdeclare{techreport}{Stahl1994}
\bibitem{Stahl1994}
\bibinfo{author}{Robert~J. \surnamestart Stahl\surnameend}
  (\bibinfo{year}{1994}): \emph{\bibinfo{title}{Using ''Think-Time'' and
  ''Wait-Time'' Skillfully in the Classroom}}.
\newblock \bibinfo{type}{Technical Report}, \bibinfo{institution}{ERIC Digest}.
\newblock \urlprefix\url{http://files.eric.ed.gov/fulltext/ED370885.pdf}.

\bibitemdeclare{inproceedings}{Stojanovic2011a}
\bibitem{Stojanovic2011a}
\bibinfo{author}{Sana \surnamestart Stojanovi{\'c}\surnameend},
  \bibinfo{author}{Vesna \surnamestart Pavlovi{\'c}\surnameend} \&
  \bibinfo{author}{Predrag \surnamestart Jani{\v{c}}i{\'c}\surnameend}
  (\bibinfo{year}{2011}): \emph{\bibinfo{title}{A {C}oherent {L}ogic {B}ased
  {G}eometry {T}heorem {P}rover {C}apable of {P}roducing {F}ormal and
  {R}eadable {P}roofs}}.
\newblock In \bibinfo{editor}{Pascal \surnamestart Schreck\surnameend},
  \bibinfo{editor}{Julien \surnamestart Narboux\surnameend} \&
  \bibinfo{editor}{J{\"u}rgen \surnamestart Richter-Gebert\surnameend},
  editors: {\sl \bibinfo{booktitle}{Automated Deduction in Geometry: 8th
  International Workshop, ADG~2010, Munich, Germany, July~22-24, 2010, Revised
  Selected Papers}}, {\sl \bibinfo{series}{Lecture Notes in Artificial
  Intelligence}} \bibinfo{volume}{6877}, \bibinfo{publisher}{Springer}, pp.
  \bibinfo{pages}{201--220}, \doi{10.1007/978-3-642-25070-5\_12}.

\bibitemdeclare{article}{Sutcliffe2018}
\bibitem{Sutcliffe2018}
\bibinfo{author}{Geoffrey \surnamestart Sutcliffe\surnameend}
  (\bibinfo{year}{2016}): \emph{\bibinfo{title}{The 8th IJCAR automated theorem
  proving system competition - CASC-J8}}.
\newblock {\sl \bibinfo{journal}{AI Communications}}
  \bibinfo{volume}{29}(\bibinfo{number}{5}), pp. \bibinfo{pages}{607--619},
  \doi{10.3233/AIC-160709}.

\bibitemdeclare{inproceedings}{Wang95}
\bibitem{Wang95}
\bibinfo{author}{D.~\surnamestart Wang\surnameend} (\bibinfo{year}{1995}):
  \emph{\bibinfo{title}{Reasoning about geometric problems using an elimination
  method}}.
\newblock In \bibinfo{editor}{J.~\surnamestart Pfalzgraf\surnameend} \&
  \bibinfo{editor}{D.~\surnamestart Wang\surnameend}, editors: {\sl
  \bibinfo{booktitle}{Automated Pratical Reasoning}},
  \bibinfo{publisher}{Springer}, \bibinfo{address}{New York}, pp.
  \bibinfo{pages}{147--185}, \doi{10.1007/978-3-7091-6604-8\_8}.

\bibitemdeclare{misc}{Wiedijk2000}
\bibitem{Wiedijk2000}
\bibinfo{author}{Freek \surnamestart Wiedijk\surnameend}
  (\bibinfo{year}{2000}): \emph{\bibinfo{title}{The {de Bruijn} factor}}.
\newblock \bibinfo{howpublished}{Poster at International Conference on Theorem
  Proving in Higher Order Logics (TPHOL2000)}.
\newblock \bibinfo{note}{Portland, Oregon, USA, 14-18 August 2000}.

\bibitemdeclare{inbook}{Wu1984}
\bibitem{Wu1984}
\bibinfo{author}{W.T. \surnamestart Wu\surnameend} (\bibinfo{year}{1984}):
  \emph{\bibinfo{title}{Automated Theorem Proving: After 25 Years}}, chapter
  \bibinfo{chapter}{On the decision problem and the mechanization of theorem
  proving in elementary geometry}, pp. \bibinfo{pages}{213--234}.
\newblock \bibinfo{volume}{29}, \bibinfo{publisher}{American Mathematical
  Society}, \doi{10.1090/conm/029}.


\bibitemdeclare{article}{ye2010a}
\bibitem{ye2010a}
\bibinfo{author}{Zheng \surnamestart Ye\surnameend},
  \bibinfo{author}{Shang-Ching \surnamestart Chou\surnameend} \&
  \bibinfo{author}{Xiao-Shan \surnamestart Gao\surnameend}
  (\bibinfo{year}{2010}): \emph{\bibinfo{title}{Visually {D}ynamic
  {P}resentation of {P}roofs in {P}lane {G}eometry: {P}art 1. {B}asic
  {F}eatures and the {M}anual {I}nput {M}ethod}}.
\newblock {\sl \bibinfo{journal}{Journal of Automated Reasoning}}
  \bibinfo{volume}{45}(\bibinfo{number}{3}), pp. \bibinfo{pages}{213--241},
  \doi{10.1007/s10817-009-9162-5}.

\bibitemdeclare{article}{ye2010b}
\bibitem{ye2010b}
\bibinfo{author}{Zheng \surnamestart Ye\surnameend},
  \bibinfo{author}{Shang-Ching \surnamestart Chou\surnameend} \&
  \bibinfo{author}{Xiao-Shan \surnamestart Gao\surnameend}
  (\bibinfo{year}{2010}): \emph{\bibinfo{title}{Visually {D}ynamic
  {P}resentation of {P}roofs in {P}lane {G}eometry: {P}art 2. {A}utomated
  {G}eneration of {V}isually {D}ynamic {P}resentations with the {F}ull-{A}ngle
  {M}ethod and the {D}eductive {D}atabase {M}ethod}}.
\newblock {\sl \bibinfo{journal}{Journal of Automated Reasoning}}
  \bibinfo{volume}{45}(\bibinfo{number}{3}), pp. \bibinfo{pages}{243--266},
  \doi{10.1007/s10817-009-9163-4}.

\bibitemdeclare{incollection}{Ye2011}
\bibitem{Ye2011}
\bibinfo{author}{Zheng \surnamestart Ye\surnameend},
  \bibinfo{author}{Shang-Ching \surnamestart Chou\surnameend} \&
  \bibinfo{author}{Xiao-Shan \surnamestart Gao\surnameend}
  (\bibinfo{year}{2011}): \emph{\bibinfo{title}{An Introduction to Java
  Geometry Expert}}.
\newblock In \bibinfo{editor}{Thomas \surnamestart Sturm\surnameend} \&
  \bibinfo{editor}{Christoph \surnamestart Zengler\surnameend}, editors: {\sl
  \bibinfo{booktitle}{Automated Deduction in Geometry}}, {\sl
  \bibinfo{series}{Lecture Notes in Computer Science}} \bibinfo{volume}{6301},
  \bibinfo{publisher}{Springer Berlin Heidelberg}, pp.
  \bibinfo{pages}{189--195}, \doi{10.1007/978-3-642-21046-4\_10}.

\bibitemdeclare{article}{zhang1995a}
\bibitem{zhang1995a}
\bibinfo{author}{Jing-Zhong \surnamestart Zhang\surnameend},
  \bibinfo{author}{Shang-Ching \surnamestart Chou\surnameend} \&
  \bibinfo{author}{Xiao-Shan \surnamestart Gao\surnameend}
  (\bibinfo{year}{1995}): \emph{\bibinfo{title}{Automated production of
  traditional proofs for theorems in {E}uclidean geometry: {I}. {T}he {H}ilbert
  intersection point theorems}}.
\newblock {\sl \bibinfo{journal}{Annals of Mathematics and Artificial
  Intelligence}} \bibinfo{volume}{13}(\bibinfo{number}{1--2}), pp.
  \bibinfo{pages}{109--137}, \doi{10.1007/BF01531326}.

\end{thebibliography}

\end{document}